\documentclass[aps,preprintnumbers,amsmath,amssymb,nofootinbib,superscriptaddress,notitlepage,10pt]{revtex4-1}
\usepackage[utf8]{inputenc}
\usepackage[sort&compress]{natbib}
\usepackage{comment}
\usepackage{amssymb}
\usepackage{color}
\usepackage{rotating}
\usepackage{amsmath}
\usepackage{ulem}
\usepackage{overpic}
\usepackage{graphicx}
\usepackage{epstopdf}
\usepackage{dcolumn}
\usepackage{bm}
\usepackage{chngpage}
\usepackage{multirow}
\usepackage{slashed}
\usepackage{indentfirst}
\usepackage{booktabs}
\usepackage{amssymb,amsfonts}
\usepackage{amsmath}
\usepackage{color}
\usepackage{hyperref}

\begin{document}

\title{Quadruply charmed baryons as heavy quark symmetry partners of the $D_{s0}^*(2317)$}

\author{Tian-Wei Wu}

\author{Ming-Zhu Liu}
\affiliation{School of Physics,
  Beihang University, Beijing 100191, China}

\author{Li-Sheng Geng}\email{lisheng.geng@buaa.edu.cn}
\affiliation{School of Physics, Beihang University, Beijing 100191, China}
\affiliation{
Beijing Key Laboratory of Advanced Nuclear Materials and Physics,
Beihang University, Beijing 100191, China}
\affiliation{School of Physics and Microelectronics, Zhengzhou University, Zhengzhou, Henan 450001, China}
\affiliation{Beijing Advanced Innovation Center for Big Data-Based Precision Medicine, School of Medicine and Engineering, Beihang University, Beijing, 100191}

\author{Emiko Hiyama}\email{hiyama@riken.jp}
\affiliation{Department of Physics, Kyushu University, Fukuoka 819-0395}
\affiliation{Japan, RIKEN Nishina Center, RIKEN, Wako 351-0198, Japan}

\author{Manuel Pavon Valderrama}\email{mpavon@buaa.edu.cn}
\affiliation{School of Physics,
Beihang University, Beijing 100191, China}

\author{Wen-Ling Wang}\email{wangwenling@buaa.edu.cn}
\affiliation{School of Physics,
Beihang University, Beijing 100191, China}

\date{\today}


\begin{abstract}
Both unitary chiral theories
and lattice QCD simulations show that the $DK$ interaction is attractive and can
form a bound state, namely,  $D^*_{s0}(2317)$.
 Assuming the validity of the heavy antiquark-diquark symmetry (HADS),
the $\Xi_{cc}\bar{K}$ interaction is  the same as the $DK$ interaction,
which implies the existence of a $\Xi_{cc}\bar{K}$ bound state with a binding energy of $49-64$ MeV. In this work, we  study whether  a $\Xi_{cc}\Xi_{cc}\bar{K}$ three-body  system binds.
The $\Xi_{cc}\Xi_{cc}$  interaction is described by exchanging $\pi$, $\sigma$, $\rho$, and $\omega$ mesons, with the corresponding couplings related to those of the $NN$  interaction via the quark model.
 We indeed find a $\Xi_{cc}\Xi_{cc}\bar{K}$  bound state, with  quantum numbers $J^P=0^-$, $I=\frac{1}{2}$, $S=1$ and $C=4$, and a
 binding energy of $80-118$ MeV.  It is interesting
to note  that this system is very similar to the well-known $NN\bar{K}$ system, which has been studied extensively
both  theoretically and experimentally. Within the same framework, we show the existence of
a  $NN\bar{K}$ state with a binding energy of $35-43$ MeV, consistent with the results of
other theoretical works and experimental data, which serves as a consistency check on the predicted
$\Xi_{cc}\Xi_{cc}\bar{K}$ bound state.
\end{abstract}

\maketitle

\section{Introduction}
In 2007, the LHCb Collaboration reported the observation of
a doubly charmed baryon, the $\Xi_{cc}^{++}$~\cite{Aaij:2017ueg}.
Its quark content is $ccu$, where it is interesting to notice that
we expect the $cc$ charmed quark pair to be tightly packed together.
The theoretical reason behind this is heavy antiquark-diquark symmetry
(HADS)~\cite{Savage:1990di}, a type of heavy-quark symmetry stating
that a heavy-quark pair behaves approximately as a heavy antiquark.
In practical terms what this means is that the structure of
the doubly heavy baryon is the same as the one of a heavy antimeson,
i.e. the wave function of the light-quark within
the $\Xi_{cc}^{(*)}$ baryon is the same as that in the $\bar{D}^{(*)}$ meson
(modulo corrections owing to the finite charm quark mass).
Consequently, HADS also implies that many of the findings related to
${D}^{(*)}$ mesons are likely to apply to $\Xi_{cc}^{(*)}$ baryons.
For instance, if there are
$D K$~\cite{Barnes:2003dj,Kolomeitsev:2003ac,Hofmann:2003je,vanBeveren:2003kd}
and
$D D K$~\cite{SanchezSanchez:2017xtl,MartinezTorres:2018zbl,Wu:2019vsy}
bound states the same is expected to happen to the $\Xi_{cc} \bar{K}$
and $\Xi_{cc} \Xi_{cc} \bar{K}$ systems.
This last system will be the subject of the present manuscript.


The existence of a $D K$ bound state is usually argued on the basis of
the experimental location of the
$D_{s0}^*(2317)$~\cite{Aubert:2003fg,Besson:2003cp,Krokovny:2003zq}.
The mass of this resonance is excessively low to be accommodated as a
$c {\bar s}$ state in the quark-model.
Yet from chiral symmetry we expect the $DK$ interaction to be
really strong and attractive, leading to the natural explanation
that the $D_{s0}^*(2317)$ is a bound state~\cite{Barnes:2003dj,Kolomeitsev:2003ac,Hofmann:2003je,vanBeveren:2003kd}.
Indeed the attraction in the $DK$ system has been repeatedly shown
to be strong enough to form this state \cite{Mohler:2013rwa, Altenbuchinger:2013vwa, Torres:2014vna,Liu:2012zya,Lang:2014yfa,Bali:2017pdv}.
Now, if we consider HADS, then the binding of the $DK$ system implies that
 the $\Xi_{cc}\bar{K}$ system should bind too,
a conclusion which  has been pointed out
in a series of theoretical works.
For instance, Ref.~\cite{Guo:2011dd} predicts an isoscalar
$\Xi_{cc}\bar{K}$ bound state at about $60 \pm 20\,{\rm MeV}$ below threshold.
Ref.~\cite{Guo:2017vcf} considers the $\Xi_{cc}\bar{K}$-$\Omega_{cc}\eta$ coupled
system, for which binding happens at about $150\,{\rm MeV}$ below threshold.
In Ref.~\cite{Meng:2018zbl} the authors calculated the $\Xi_{cc}\bar{K}$
scattering length to be $2.15\,{\rm fm}$, which being positive
(and provided that the system is attractive) indicates
the existence of a bound state.
In Ref.~\cite{Yan:2018zdt}, in addition to  the next-to-leading order chiral potentials,  the $P$-wave excitation between the two heavy quarks was taken into account as a dynamical degree of freedom,
a  $\Xi_{cc}\bar{K}$ bound state
with a binding energy of 50 MeV was predicted.

The bottom-line is that the existence of a $\Xi_{cc}\bar{K}$ bound state
is really likely, at least if HADS breaking is not too large.
This immediately raises the intriguing question of
whether there exists a $\Xi_{cc}\Xi_{cc}\bar{K}$ trimer.
The reasons why such a trimer is probable are HADS and
the theoretical predictions of a $DDK$ bound state~\cite{SanchezSanchez:2017xtl,MartinezTorres:2018zbl,Wu:2019vsy,Huang:2019qmw}.
It is also interesting to notice that the probable mechanism responsible
for the formation of the $DDK$ and $\Xi_{cc}\Xi_{cc}\bar{K}$ trimers is
the strong Weinberg-Tomozawa term in the $DK$ and $\Xi_{cc} \bar{K}$
subsystems.
This feature is shared with the $N\bar{K}$ system, which is usually thought
to be the most important component of the $\Lambda(1405)$ wave function
and which also leads to the formation of $NN\bar{K}$ bound states,
as has been extensively studied in both theory~\cite{Yamazaki:2002uh,Magas:2006fn,Shevchenko:2006xy,Shevchenko:2007ke,Ikeda:2007nz,Yamazaki:2007cs,Arai:2007qj,Nishikawa:2007ex,Dote:2008in,Dote:2008hw,Ikeda:2008ub,Wycech:2008wf,Ikeda:2010tk,Uchino:2011jt,Barnea:2012qa,Bayar:2012hn,Dote:2014via,Ohnishi:2017uni,Dote:2017veg}
and experiment~\cite{Agnello:2005qj,Yamazaki:2010mu,Ichikawa:2014ydh,Agakishiev:2014dha,Ajimura:2018iyx,Sada:2016nkb,Yamazaki:2008hm,Tokiyasu:2013mwa},
with the later supporting the existence of this trimer.
Motivated by these facts, in this manuscript we will explore
the three-body $\Xi_{cc}\Xi_{cc}\bar{K}$ system,
which we conclude to be likely to bind.

The article is organized as follows: in Sec.~\ref{sec:Interactions},
we explain the $\Xi_{cc}\bar{K}$ and $\Xi_{cc}\Xi_{cc}$ interactions.
In Sec.~\ref{sec:GEM}  we construct the three-body wave functions and
solve the corresponding Schr\"odinger equation
for the $\Xi_{cc}\Xi_{cc}\bar{K}$ system using the Gaussian Expansion Method
(GEM).
In Sec.~\ref{sec:Results}, we present our predictions and discuss
the theoretical uncertainties associated with them.
Finally, we summarize our results in Sec.~\ref{sec:Summary}.

\section{Two-body Interactions}
\label{sec:Interactions}

Here we will explain in detail how to derive the $\Xi_{cc}\bar{K}$
and $\Xi_{cc}\Xi_{cc}$ interactions.
For the case of the $\Xi_{cc}\bar{K}$ system the most important part of
the interaction is given by the Weinberg-Tomozawa term,
which happens to be identical to that of the $D K$ system
owing to HADS and chiral symmetry.
For the $\Xi_{cc} \Xi_{cc}$ system we will resort to
the one boson exchange (OBE) model, where the non-relativistic potential
between the two baryons is determined by the exchange of a few light mesons
(the pion, the sigma, the rho and the omega).
For determining the coupling constants of
these light mesons to the doubly-charmed baryons
we will resort to HADS and the information we can deduce
from the $DD$ two charmed-meson system
and the quark model.

\subsection{0($\frac{1}{2}^-$) $\Xi_{cc}\bar{K}$ potential}

The most important contribution to the $\Xi_{cc}\bar{K}$ interaction
is the Weinberg-Tomozawa term between the kaon and the doubly charmed baryon,
which in a non-relativistic normalization reads
\begin{eqnarray}
  V_{WT}(\vec{q}) = - \frac{C_{WT}(I)}{2 f_{\pi}^2} \, ,
\end{eqnarray}
with $f_{\pi} \simeq 130\,{\rm MeV}$ and $C_{WT}(0) = 2$, $C_{WT}(1) = 0$
for the isoscalar and isovector channels, respectively.
This coincides with the standard half-relativistic (relativistic kaon and
non-relativistic baryon) normalization
$C_{WT}(I) (\omega_K + \omega_K') / 2f_{\pi}^2$.
This potential is exactly the same one as for the $DK$ system as a consequence
of two independent facts: the universality of the WT term (and the fact that
$D$ and $\Xi_{cc}$ belong to the $\bar{3}$ and $\bar{3}$ representations
of SU(3)-flavor) and HADS, which also implies the same interaction
in both systems.

The Fourier-transform of the previous potential in coordinate space is
\begin{eqnarray}
  V_{WT}(\vec{r}) = - \frac{C_{WT}(I)}{2 f_{\pi}^2}\,\delta^{(3)}(\vec{r}) \, ,
  \label{eq:V-WT}
\end{eqnarray}
which is singular and requires regularization.
For that purpose we will choose a Gaussian regulator of the type
\begin{eqnarray}
  V_{WT}(\vec{r}) = - \frac{C_{WT}(I)}{2 f_{\pi}^2}\,
  \frac{e^{-(r/R_c)^2}}{\pi^{3/2} R_c^3} \, ,
\end{eqnarray}
where $R_c$ is a coordinate space cutoff.
However the previous expression is still problematic, as the prediction
of a bound $\Xi_{cc} \bar{K}$ state and its binding energy depends
on the cutoff.
If there is an experimentally known bound state,
then it is easy to choose the cutoff in order to reproduce that bound state.
Though this is not the case for the $\Xi_{cc} \bar{K}$, it happens that the
$D_{s0}^*(2317)$ is suspected to be a $D K$ bound state.
From this we can set the cutoff in the $D K$ system, which, owing to HADS,
should be the same cutoff as in the $\Xi_{cc} \bar{K}$ system.

But there is the more powerful approach of fully renormalizing
the $\Xi_{cc} \bar{K}$ / $D K$ interaction,
which is what we will do here.
For that, we allow the strength of the WT term to vary with the cutoff
\begin{eqnarray}
  V(\vec{r}) = C(R_c)\,\frac{e^{-(r/R_c)^2}}{\pi^{3/2} R_c^3} \, ,
\end{eqnarray}
where for every value of $R_c$ we determine $C(R_c)$ by reproducing
the $D_{s0}^*(2317)$ as a $DK$ bound state.
After this we can predict the binding energy of the $\Xi_{cc} \bar{K}$ system.~\footnote{
HADS is known  to be broken at the level of  $\Lambda_\mathrm{QCD}/(m_Q\nu)$~\cite{Savage:1990di},  where
$\nu$ is the velocity of the heavy quark pair. For a charm quark pair, $m_Q\nu\sim 0.8$ GeV~\cite{Hu:2005gf}, we obtain a breaking of  $25-40\%$ for HADS. At this moment, there is no concrete experimental information that can help
us to estimate the level of the breaking of HADS. Future discovery of the spin $3/2$ partner of the $\Xi_{cc}$
should give us a clue, since HADS states
\begin{equation}
m_{\Xi_{cc}^*}-m_{\Xi_{cc}}=\frac{3}{4}(m_{D^*}-m_D)\approx106.5\,\mathrm{MeV}.
\end{equation}
Lattice QCD studies~\cite{Lewis:2001iz, Mathur:2002ce, Flynn:2003vz,Brown:2014ena,Padmanath:2015jea} and various models~\cite{Karliner:2014gca,Weng:2018mmf,Yu:2018com,Roberts:2007ni,Anselmino:1987vk,Kroll:1988cd,Chang:2003ua,Richard:1994ae}
suggest a breaking up to  $40\%$. As a result, the same as Ref.~\cite{Liu:2018zzu}, we study the uncertainties due to  the breaking of HADS by varying the couplings, $C(R_c)$ and $C_S$,  from their central values by $25\%$.}
In addition to this, we can also modify the previous potential
with a shorter-range contribution as follows
\begin{eqnarray}
  V_{WT}(\vec{r}) &=& C(R_c)\,\frac{e^{-(r/R_c)^2}}{\pi^{3/2} R_c^3} +
  C_S\,\frac{e^{-(r/R_S)^2}}{\pi^{3/2} R_S^3} \nonumber \\
  &=& C_L'\,e^{-(r/R_c)^2} + C_S'\,e^{-(r/R_S)^2}
  \, ,
  \label{eq:V2-final}
\end{eqnarray}
where we take $R_S < R_c$ and $C_S > 0$ (i.e. repulsive) with $|C_S| > |C(R_c)|$.
Notice that we also define the couplings $C_L'$ and $C_L'$, which are equivalent
to $C(R_c)$ and $C_S$ but more convenient to use.
The purpose of this modification is to take into account the fact that the
subleading order corrections to the WT term for the $DK$ interaction are
repulsive in nature (with the same thing happening in the
$\Xi_{cc} \bar{K}$ system owing to HADS)~\cite{Altenbuchinger:2013vwa}.
For concreteness we will take $R_c = 0.5-2.0\,{\rm fm}$
and $R_S = 0.1\,{\rm fm}$.
For $C_S'$ we will consider two possibilities: $C_S' = 0$ (i.e. we ignore
the existence of subleading order corrections) and
${C}_S' = 1\,{\rm GeV}$ (to exaggerate the strength
of these corrections).
The binding energy we predict for the $\Xi_{cc} \bar{K}$ state lies
in the range $B_2 = (50-60)\,{\rm MeV}$ and are almost
independent of  $\hat{C}_S$.
Concrete results for each cutoff can be checked in Table \ref{Results:XiccXiccKbar},
which we will discuss later on.

In addition to the strong force, the $\Xi_{cc} \bar{K}$ system also receives
contributions from the electromagnetic force if the antikaon happens
to be charged (the $\Xi_{cc}$ is always charged, with its particle
states being $\Xi_{cc}^+$ or $\Xi_{cc}^{++}$).
The two particle components of the $I=0$ channel in which the $\Xi_{cc} \bar{K}$
interaction is expected to be stronger are
\begin{eqnarray}
  | \Xi_{cc} \bar{K} (I=0) \rangle = \frac{1}{\sqrt{2}}\,
  \left[
    | \Xi_{cc}^{+} \bar{K}^0 \rangle + | \Xi_{cc}^{++} \bar{K}^- \rangle
    \right] \, .
\end{eqnarray}
For each of these components the Coulomb force reads
\begin{eqnarray}
  V^C_{\Xi_{cc}^{+} \bar{K}^{0}}(r) &=& 0 \, , \\
  V^C_{\Xi_{cc}^{++} K^{-}}(r) &=& -2 \frac{\alpha}{r} \, ,
\end{eqnarray}
or, equivalently, if we write it with isospin operators
\begin{equation}
  V^C_{\Xi_{cc} \bar{K}}(r)= \frac{\alpha}{r}\,\frac{(\tau_{z,1} +3)}{2}\,
  \frac{(-1+\tau_{z,2})}{2} \, ,
\end{equation}
with $\tau_{z,1}$ and $\tau_{z,2}$ the third component of the isospin operator
for the doubly charmed baryon and antikaon, respectively.
The problem is that the Coulomb force breaks isospin symmetry,
which can be dealt with in two ways.
The simplest one is to average the Coulomb force over the particle components
of the $I=0$ state
\begin{eqnarray}
  \langle \Xi_{cc} \bar{K} (I=0) | V^C(r) | \Xi_{cc} \bar{K} (I=0) \rangle =
  - \frac{\alpha}{r} \, . \label{eq:Coulomb-opt-A}
\end{eqnarray}
The other possibility is to consider the $| \Xi_{cc}^{+} \bar{K}^0 \rangle$
and $| \Xi_{cc}^{++} \bar{K}^- \rangle$ components separately as channels
1 and 2, in which case we can write the full potential as
\begin{eqnarray}
  V(r) = V_{WT}(r)\,\frac{1}{2}
  \begin{pmatrix}
    1 & 1 \\
    1 & 1
  \end{pmatrix} -
  \frac{\alpha}{r}
  \begin{pmatrix}
    0 & 0 \\
    0 & 2
  \end{pmatrix} \, . \label{eq:Coulomb-opt-B}
\end{eqnarray}
In this work we will choose the first option, the one described
by Eq.~(\ref{eq:Coulomb-opt-A}), as this will greatly
simplify the formalism required to do the calculations,
particularly once we consider the three-body system.

\subsection{The $\Xi_{cc}\Xi_{cc}$ potential in the one boson exchange model}

In the OBE model the interaction between two hadrons is described in terms of
the exchange of a series of light mesons, which most commonly include
the pion ($\pi$), sigma ($\sigma$), rho ($\rho$) and omega ($\omega$),
but sometimes a few more bosons in its more sophisticated incarnations.
Indeed the OBE model has provided one of the most quantitatively successful
description of the nuclear forces~\cite{Machleidt:1987hj,Machleidt:1989tm}
and in principle there is nothing impeding its application to other
two-hadron systems.
Regarding hadronic molecules, it is interesting to notice that the original
speculations about their existence were based
on the OBE model~\cite{Voloshin:1976ap},
which later has been widely used for predicting or explaining
molecular states~\cite{Liu:2008fh, Liu:2007bf, Yang:2011wz, Sun:2011uh,Chen:2017jjn,Yamaguchi:2019vea}.

The particular version of the OBE model we will use is the one developed
in Ref.~\cite{Liu:2019stu} for the $D\bar{D}$ and $DD$
family of charmed meson molecules,
which in turn can be related to the $\Xi_{cc} \Xi_{cc}$ case via HADS.
The most important difference of Ref.~\cite{Liu:2019stu} with previous
implementations of the OBE model for heavy hadron molecules
is the inclusion of a few of the ideas of the {\it renormalized OBE model}
of Ref.~\cite{Cordon:2009pj}.
In particular we partially renormalize the OBE model, by which we mean
the following: the OBE model contains a form factor and a cutoff
$\Lambda$, where we determine the cutoff $\Lambda$ from
the condition of reproducing a known molecular state.
The molecular state chosen is the $X(3872)$, of which there is evidence that
it might be a $D^*\bar{D}$ molecule with $J^{PC} = 1^{++}$ and isospin $I=0$.
For the case of a monopolar form factor the resulting  cutoff is
$\Lambda_X = 1.01^{+0.19}_{-0.10}\,{\rm GeV}$.

We will not explain in detail here how to derive the OBE potential
for the general $\Xi_{cc}^{(*)} \Xi_{cc}^{(*)}$ system from HADS and
the potential developed in Ref.~\cite{Liu:2019stu}
for the $D^{(*)} D^{(*)}$ system.
Instead we will simply indicate how to do it, where the starting point are
the definitions of the charmed meson and doubly charmed baryon superfields
\begin{eqnarray}
  H_c &=& \frac{1}{\sqrt{2}}\left[ D + \vec{\sigma} \cdot \vec{D}^* \right]
  \quad , \quad
  \vec{T}_{cc} = \frac{1}{\sqrt{3}} \vec{\sigma}\,\Xi_{cc} + \vec{\Xi}^*_{cc} \, ,
\end{eqnarray}
which group the $D$, $D^*$ meson ($\Xi_{cc}$, $\Xi_{cc}^*$ baryon) fields
into a single superfield with good properties with respect to
rotations of the heavy quark spin.
For implementing HADS there are several possibilities, of which
we briefly explain two.
One is to group the two superfields $H_c$ and $T_{cc}$ into a new superfield
$\mathcal{H}_c$ which is invariant under HADS.
The other is to simply notice that the previous procedure at the end amount
to make the following substitutions in the Lagrangians:
\begin{eqnarray}
  {\rm Tr}\left[ H_c^\dagger \mathcal{O} H_c \right]
  \to T_{cc}^{\dagger} \, \mathcal{O} \, T_{cc} \, ,
\end{eqnarray}
with $\mathcal{O}$ some arbitrary spin operator acting on the superfields.
If we do this with the OBE Lagrangian of Ref.~\cite{Liu:2019stu}
we will be able to derive the potentials we will write down below.

The outcome of the previous procedure for the particular case of
the $\Xi_{cc} \Xi_{cc}$ system is
\begin{eqnarray}
  V_{\rm OBE} &=& V_{\pi} + V_{\sigma} + V_{\rho} + V_{\omega} \, ,
    \label{eq:V-OBE-schema}
\end{eqnarray}
where the contributions from the $\pi$, $\sigma$, $\rho$ and $\omega$ read
\begin{eqnarray}
  V_{\pi}(\vec{r}) &=&
  \vec{\tau}_1 \cdot \vec{\tau}_2\,\frac{g^2}{6 f_{\pi}^2}\,\Big[
    - \frac{1}{9}\,\vec{\sigma}_1 \cdot \vec{\sigma}_2\,m_{\pi}^3\,
    d(m_\pi r, \frac{\Lambda}{m_{\pi}})
    \nonumber \\ && \quad
    + \frac{1}{9}\,\vec{\sigma}_1 \cdot \vec{\sigma}_2\,m_{\pi}^3\,
    W_Y(m_{\pi} r, \frac{\Lambda}{m_{\pi}})
    \nonumber \\ && \quad
    + \frac{1}{9}\,S_{12}(\vec{r})\,m_{\pi}^3\,
    W_T(m_{\pi} r, \frac{\Lambda}{m_{\pi}}) \Big] \, , \label{eq:V-OBE-pi} \\
  V_{\sigma}(\vec{r}) &=& -{g_{\sigma}^2}\,m_{\sigma}\,
  W_Y(m_{\sigma} r, \frac{\Lambda}{m_{\sigma}})
  \, , \\
  V_{\rho}(\vec{r}) &=& \vec{\tau}_1 \cdot \vec{\tau}_2\,\Big[
    {g_{\rho}^2}\,m_{\rho}\,W_Y(m_{\rho} r, \frac{\Lambda}{m_{\rho}}) \nonumber \\
    && \quad + \frac{f_{\rho}^2}{4 M^2}\,\Big(
    -\frac{2}{27}\,\vec{\sigma}_1 \cdot \vec{\sigma}_2 \, m_{\rho}^3 \,
    W_Y(m_{\rho} r, \frac{\Lambda}{m_{\rho}})
    \nonumber \\ && \qquad
    +\frac{2}{27}\,\vec{\sigma}_1 \cdot \vec{\sigma}_2 \, m_{\rho}^3 \,
    W_Y(m_{\rho} r, \frac{\Lambda}{m_{\rho}})
    \nonumber \\ && \qquad
    -\frac{1}{27}\,S_{12}(\hat{r})\, m_{\rho}^3 \,
    W_T(m_{\rho} r, \frac{\Lambda}{m_{\rho}}) \,\,
    \Big) \, \Big] \, , \\
    V_{\omega}(\vec{r}) &=&
    {g_{\omega}^2}\,m_{\omega}\,W_Y(m_{\omega} r, \frac{\Lambda}{m_{\omega}})
    \nonumber \\
    && \quad \frac{f_{\omega}^2}{4 M^2}\,\,\Big(
    -\frac{2}{27}\,\vec{\sigma}_1 \cdot \vec{\sigma}_2 \,
    m_{\omega}^3\,d(m_{\omega} r, \frac{\Lambda}{m_{\omega}})
    \nonumber \\ && \qquad
    +\frac{2}{27}\,\vec{\sigma}_1 \cdot \vec{\sigma}_2 \,
    m_{\omega}^3 \, W_Y(m_{\omega} r, \frac{\Lambda}{m_{\omega}})
    \nonumber \\ && \qquad
    -\frac{1}{27}\,S_{12}(\hat{r})\, m_{\omega}^3 \,
    W_T(m_{\omega} r, \frac{\Lambda}{m_{\omega}}) \,\,
    \Big) \, , \label{eq:V-OBE-omega}
\end{eqnarray}
where for a monopolar form factor the functions $d$, $W_Y$ and $W_T$ take
the form
\begin{eqnarray}
  d(x, \lambda) &=& \frac{(\lambda^2 - 1)^2}{2 \lambda}\,
  \frac{e^{-\lambda x}}{4 \pi} \, , \\
  W_Y(x, \lambda) &=& W_Y(x) - \lambda W_Y(\lambda x) \nonumber \\ && -
  \frac{(\lambda^2 - 1)}{2 \lambda}\,\frac{e^{-\lambda x}}{4 \pi} \, , \\
  W_T(x, \lambda) &=& W_T(x) - \lambda^3 W_T(\lambda x) \nonumber \\ && -
  \frac{(\lambda^2 - 1)}{2 \lambda}\,\lambda^2\,
  \left(1 + \frac{1}{\lambda x} \right)\,\frac{e^{-\lambda x}}{4 \pi} \, .
\end{eqnarray}
For the coupling constants we follow Ref.~\cite{Liu:2019stu} and take
$g = 0.60$, $g_{\sigma} = 3.4$, $g_{\rho} = g_{\omega} = 2.6$,
$f_{\rho} = f_{\omega} = g_{\omega} \kappa_{\omega}$,
$\kappa_{\omega} = 4.5$ and
$M = 1867\,{\rm MeV}$.

Finally we have to include the Coulomb piece,
which written in the isospin basis reads
\begin{equation}
  V^C_{\Xi_{cc} \Xi_{cc}}(r)=
  \frac{\alpha}{r}\,\frac{(\tau_{z,1} +3)}{2}\,\frac{(\tau_{z,2} +3)}{2} \, .
\end{equation}
If we consider the $S=0$ (singlet) S-wave $\Xi_{cc} \Xi_{cc}$ molecule,
there are three possible isospin states
corresponding to the $\Xi_{cc}^+ \Xi_{cc}^+$, $\Xi_{cc}^+ \Xi_{cc}^{++}$
and $\Xi_{cc}^{++} \Xi_{cc}^{++}$ systems.
If we consider the $S=1$ (triplet) case, this corresponds to
$\Xi_{cc}^+ \Xi_{cc}^{++}$ and a Coulomb potential
\begin{equation}
  V^C_{\Xi_{cc} \Xi_{cc}}(r; S=1, I = 0) =
  2\,\frac{\alpha}{r} \, ,
\end{equation}

Concrete calculations with the previous parameters indicate that there is
no singlet bound state but that a triplet bound state --- the charming
deuteron --- will bind for $\Lambda_S \geq 994\,{\rm MeV}$ without
Coulomb and $\Lambda_C \geq 1112\,{\rm MeV}$ with Coulomb, consistent with the previous
prediction of Ref.~\cite{Valderrama:2019sid}.


\section{Gaussian expansion method}\label{sec:GEM}

\begin{figure}[!h]
  \centering
  \includegraphics[width=10cm]{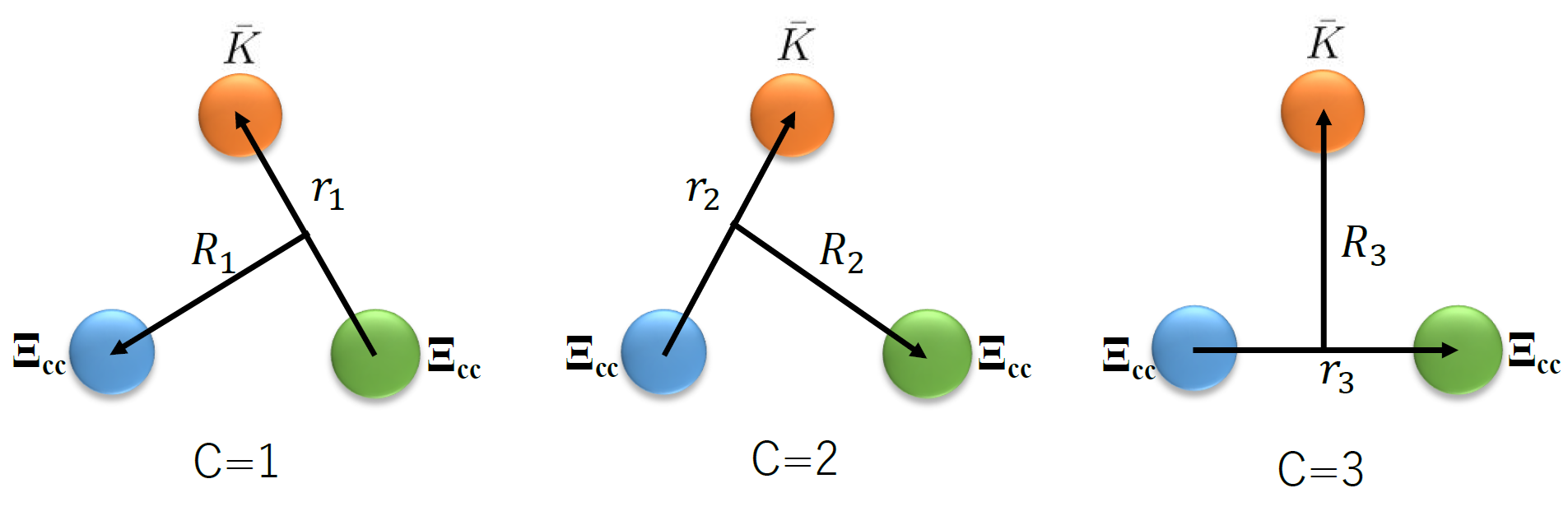}\\
  \caption{Three permutations of the Jacobi coordinates for the $\Xi_{cc}\Xi_{cc}\bar{K}$ system.}\label{Jac}
\end{figure}

Once we have determined all the relevant two-body interactions,
we are ready to explore the $\Xi_{cc}\Xi_{cc}\bar{K}$ three-body system.
For this we will use the  Gaussian Expansion Method (GEM) \cite{Kamimura:1988zz,Hiyama:2003cu},
which is an efficient method to solve few-body systems.
The starting point is the Sch\"{o}dinger equation
\begin{equation}\label{schd}
  H\Psi_{JM}^{total}=E\Psi_{JM}^{total},
\end{equation}
with the Hamiltonian
\begin{equation}\label{hami}
  \hat{H} = \sum_{i=1}^{3} \frac{p_i^2}{2m_i}-T_{c.m.} +
   V_{\Xi_{cc}\bar{K}}(r_1) + V_{\Xi_{cc}\bar{K}}(r_2) + V_{\Xi_{cc}\Xi_{cc}}(r_3),
\end{equation}
where $T_{c.m.}$ is the kinetic energy of the center of mass and $V(r)$ is
the potential between the two relevant particles.
The three possible permutations of the Jacobi coordinates
for the $\Xi_{cc}\Xi_{cc}\bar{K}$ system are
depicted in Fig.\ref{Jac}.
The total wave function can be expressed as the sum of the amplitudes of
the three re-arrangement channels ($c = 1-3$), i.e. the permutations
shown in Fig.\ref{Jac}, which we write as
  \begin{equation}\label{Sch}
    \Psi_{JM}^{total}=
    \sum_{c,\alpha}C_{c,\alpha}\Psi_{JM,\alpha}^{c}(\mathbf{r}_c,\mathbf{R}_c)
  \end{equation}
  with $\mathbf{r}_c$ and $\mathbf{R}_c$ the Jacobi coordinates in channel $c$.
  As can be appreciated, the wave function is expanded in a series
  in terms of $\alpha=\{nl,NL,\Lambda,tT\}$ (explained below),
  with $C_{c,\alpha}$ the expansion coefficients.
  Here $l$ and $L$ are the orbital angular momentum
  for the coordinate $r$ and $R$,  $t$ is the isospin of
  the two-body subsystem in each channel, $\Lambda$~\footnote{This should not be confused with
  the cutoff used to regularize the OBE potential.} and $T$ are
  the total orbital angular momentum and isospin, $n$ and $N$ are
  the numbers of gaussian basis functions corresponding to coordinates
  $r$ and $R$, respectively.
  Considering that the two baryons are identical, the total wave function
  should be antisymmetric with respect to the exchange of
  the two $\Xi_{cc}$ baryons, which requires
\begin{equation}\label{exchange}
  P_{12}  \Psi_{JM}^{total}= -\Psi_{JM}^{total},
\end{equation}
where $P_{12}$ is the exchange operator of particles 1 and 2.
The wave function of  each channel has the following form
  \begin{equation}\label{dd}
    \Psi_{JM,\alpha}^{c}(\mathbf{r}_c,\mathbf{R}_c) =
    H_{T,t}^c\otimes[\Phi_{lL,\Lambda}^c(\mathbf{r}_c,\mathbf{R}_c)]_{JM}
  \end{equation}
  where $H_{T,t}^c$ is the isospin wave function, and $\Phi_{lL,\Lambda}^c$
  the orbital wave function. The isospin wave function
  in each channel reads as
  \begin{equation}\label{isospinwave}
    \begin{split}
      H_{T,t}^{c=1}& =
      [[\eta_{\frac{1}{2}}(\Xi_{cc}^2)\eta_{\frac{1}{2}}(\bar{K}^3)]_{t_1}\eta_{\frac{1}{2}}(\Xi_{cc}^1)]_{\frac{1}{2}}, \\
      H_{T,t}^{c=2}& =
      [[\eta_{\frac{1}{2}}(\Xi_{cc}^1)\eta_{\frac{1}{2}}(\bar{K}^3)]_{t_2}\eta_{\frac{1}{2}}(\Xi_{cc}^2)]_{\frac{1}{2}},\\
      H_{T,t}^{c=3}& =
      [[\eta_{\frac{1}{2}}(\Xi_{cc}^1)\eta_{\frac{1}{2}}(\Xi_{cc}^2)]_{t_3}\eta_{\frac{1}{2}}(\bar{K}^3)]_{\frac{1}{2}}.
    \end{split}
  \end{equation}
  The orbital wave function $\Phi_{lL,\Lambda}^c$ is given
  in terms of the Gaussian basis functions
    \begin{equation}\label{nj}
   \Phi_{lL,\Lambda}^c(\mathbf{r}_c,\mathbf{R}_c)=[\phi_{n_cl_c}^{G}(\mathbf{r}_c)\psi_{N_cL_c}^{G}(\mathbf{R}_c)]_{\Lambda},
  \end{equation}
   \begin{equation}\label{nj}
  \phi_{nlm}^{G}(\mathbf{r}_c)=N_{nl}r_c^le^{-\nu_n r_c^2} Y_{lm}({\hat{r}}_c),
  \end{equation}
  \begin{equation}\label{nj}
  \psi_{NLM}^{G}(\mathbf{R}_c)=N_{NL}R_c^Le^{-\lambda_n R_c^2} Y_{LM}({\hat{R}}_c).
  \end{equation}
  Here $N_{nl}(N_{NL})$ is the normalization constant of the Gaussian basis and the parameters $\nu_n$ and $\lambda_n$ are given by
  \begin{equation}\label{vn}
  \begin{split}
       \nu_n &=1/r_n^2,\qquad r_n=r_{min}a^{n-1}\quad (n=1,n_{max}), \\
       \lambda_N &=1/R_N^2,\quad R_N=R_{min}A^{N-1}\quad (N=1,N_{max}),
  \end{split}
  \end{equation}
  where $\{n_{max},r_{min},a$ or $r_{max}\}$ and  $\{N_{max},R_{min},A$ or $R_{max}\}$ are gaussian basis parameters.
\begin{table}
		\caption{Configurations of the $I(J^P)$=$\frac{1}{2}(0^-)$ $\Xi_{cc}\Xi_{cc}\bar{K}$ system together with the number of Gaussian basis used. Here, $B=\Xi_{cc}$, $\phi=\bar{K}$ and $S_{BB}$ the spin of $\Xi_{cc}\Xi_{cc}$ subsystem. Note that channel 1 and channel 2 are identical.}\label{Qumconfig}
	\begin{tabular}{c c c c c c c c c c c}
		\hline
		\hline
		$c$&$l_{B\phi}$&$L_{B\phi-B}$&$\Lambda$&$S_{BB}$&$J$&$t_{B\phi}$&$I$&$P$&$n_{max}$&$N_{max}$\\
		\hline
		1(2)&0&0&0&0&0&0&$\frac{1}{2}$&$-$&10&10\\
		1(2)&0&0&0&0&0&1&$\frac{1}{2}$&$-$&10&10\\
		\hline
		$C$&$l_{BB}$&$L_{BB-\phi}$&$\Lambda$&$S_{BB}$&$J$&$t_{BB}$&$I$&$P$&$n_{max}$&$N_{max}$\\
		\hline
		3&0&0&0&0&0&1&$\frac{1}{2}$&$-$&10&10\\
		\hline
	\end{tabular}
\end{table}

Since the $\Xi_{cc}$ is a doubly charmed $\frac{1}{2}(\frac{1}{2}^-)$ baryon and $\bar{K}$ a $\frac{1}{2}(0^-)$ meson, considering only $S$-wave interactions and the Fermi-Dirac statistics of the two identical $\Xi_{cc}$ baryons, the quantum numbers of the $\Xi_{cc}\Xi_{cc}\bar{K}$ system are $I(J^P)=\frac{1}{2}(0^-)$. All the  configurations of this three-body system are shown in Table.\ref{Qumconfig}.

As the wave function been constructed, the Sch\"{o}dinger equation of this system is transformed into a generalized matrix eigenvalue problem by the basis expansion:
\begin{equation}\label{eigenvalue problem}
[T_{\alpha \alpha'}^{ab}+V_{{\alpha \alpha'}}^{ab}-EN_{\alpha \alpha'}^{ab}]\,
C_{b,\alpha'} = 0
\, .
\end{equation}
Here, $T_{\alpha \alpha'}^{ab}$ is the kinetic matrix element, $V_{\alpha \alpha'}^{ab}$ is the potential matrix element and $N_{\alpha \alpha'}^{ab}$ is the normalization matrix element. The eigenenergy $E$ and coefficients are determined by the Rayleigh-Ritz variational principle via the gaussian basis parameters.

\section{Thomas collapse in the $\Xi_{cc} \Xi_{cc} \bar{K}$ system}
\label{sec:Thomas}

A problem with the $\Xi_{cc} \Xi_{cc} \bar{K}$ system is that it is Efimov-like.
Thus it will be possible for it to show Thomas collapse.
In principle this means that the predictions we will make will be cutoff
dependent, as the only way to stabilize the energy of the ground state
is to include a short-range repulsive three-body force.

Actually to show the existence of the Thomas collapse in this system
we will use the Efimov effect as a proxy.
The Efimov effect refers to the appearance of a geometric spectrum
in the three-body system when a few of the interacting particles
are in the unitary limit, i.e. their scattering lengths diverge.
This is complementary to the Thomas collapse: reducing the range of
the interaction is equivalent to a relative increase of
the scattering length when expressed
in units of the range.
The presence of the Efimov effect can be deduced from the Faddeev equations
for a contact-range potential, which we will not derive in detail here
but can be consulted in Refs.~\cite{Helfrich:2010yr,Valderrama:2018sap,Valderrama:2018azi}.
Instead we will simply use the results derived in other works:
if we consider a three-body system of the type $AAB$,
where $A$ and $B$ are two different species of
particles and the $AB$ interaction is resonant,
the condition for having the Efimov effect is
\begin{eqnarray}
  \lambda_{\alpha} = \frac{\sin{2 \alpha}}{2 \alpha} \leq \lambda \, ,
\end{eqnarray}
with $\lambda$ a geometric factor depending on the characteristics of
the $AB$ interaction and quantum numbers of the system, and $\alpha$
an angle given by
\begin{eqnarray}
  \alpha = {\rm arcsin}\,\left( \frac{1}{1 + \frac{m_B}{m_A}} \right) \, .
\end{eqnarray}
For the $\Xi_{cc} \Xi_{cc} \bar{K}$ system we have that
$m_A = m(\Xi_{cc})$ and $m_B = m(K)$,
which gives $\lambda_{\alpha} \simeq 0.389$.
The factor $\lambda$ is given in this case by the condition that
the $\Xi_{cc} \bar{K}$ interaction is strong
in the isospin $I = 0$ channel, and can be calculated
from the matrix element of the isospin wave functions
from the two different permutations of the doubly charmed baryons:
\begin{eqnarray}
  \lambda =
   \langle H^{c=1}_{\frac{1}{2},0} | H^{c=2}_{\frac{1}{2},0} \rangle
  = \frac{1}{2} \, .
\end{eqnarray}
From this we have that $\lambda \geq \lambda_{\alpha}$:
the conclusion is that for the $\Xi_{cc} \Xi_{cc} \bar{K}$ system the Efimov
effect can indeed happen.
But of course, owing to the fact that the $\Xi_{cc} \bar{K}$ is far
from the unitary limit (i.e. they do not form a shallow bound state),
what we expect instead is Thomas collapse.
For comparison purposes, in the $D D K$ and $N N \bar{K}$ systems
we have $\lambda_{\alpha} \simeq 0.531$ and $0.693$, respectively,
from which we deduce that there is no Thomas collapse
in these two cases.

\section{Predictions}
\label{sec:Results}

Once we have determined the required two-body inputs, the calculation of
prospective $\Xi_{cc}\Xi_{cc}\bar{K}$ trimers is straightforward.
For this we will use the GEM, which we have already explained
in Section \ref{sec:GEM}.
As already discussed, the $\Xi_{cc}\Xi_{cc}\bar{K}$ three-body system suffers
from Thomas collapse, i.e. if the range of the involved two-body
$\Xi_{cc} \bar{K}$ interaction would be reduced to zero,
the three-body system would collapse.
Of course in the real world this does not happen because the range
the $\Xi_{cc} \bar{K}$ interaction is finite, but this collapse
will manifest itself as a strong dependence on the cutoff $R_c$
we have chosen to regularize the potential.
Finally, we notice that the $\Xi_{cc}\Xi_{cc}\bar{K}$ system is completely
analogous to the $NN \bar{K}$ one: the doubly-charmed baryons and the nucleons
belong to the same irreducible representation of the spin and isospin
and are therefore interchangeable modulo two difference,
one being the masses and the other being the strength of
the WT term with the antikaon.
For this reason we will also present calculations of the $NN\bar{K}$ trimer.


\subsection{The $\frac{1}{2}(0^-)$ $\Xi_{cc}\Xi_{cc}\bar{K}$ system}

\begin{figure}
	\centering
	\includegraphics[scale=0.3]{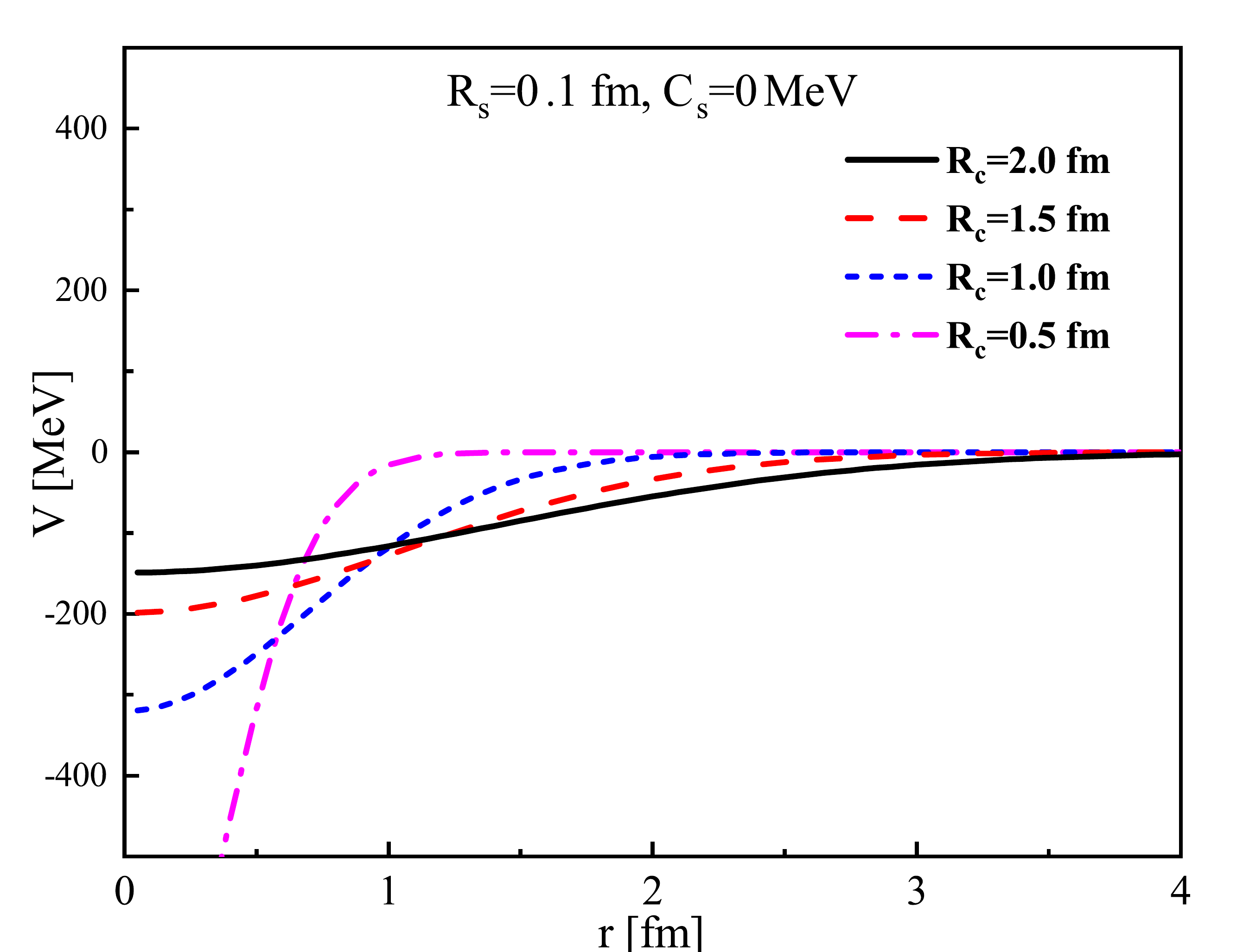}
\includegraphics[scale=0.3]{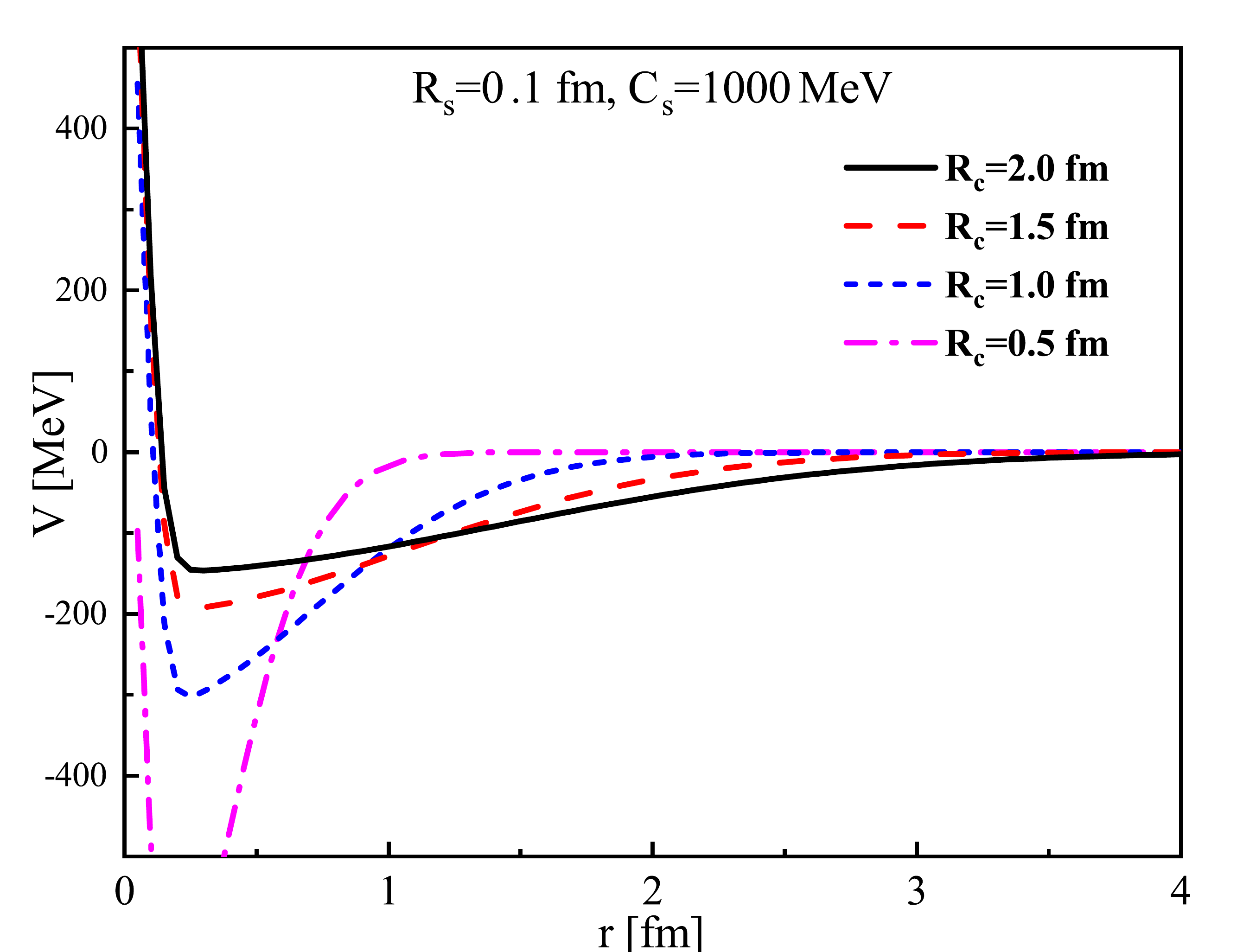}
	\caption{Weinberg-Tomozawa $\Xi_{cc}\bar{K}$ potential with specified $R_s$, $C_S$, and $R_c$. The other parameter $C(R_c)$ is determined by reproducing the $D_{s0}^*(2317)$. }\label{Fig:WTpoten}
\end{figure}

We now present here the $\Xi_{cc} \bar{K}$ dimer and
$\Xi_{cc}\Xi_{cc}\bar{K}$ trimer predictions.
We begin with the dimer, as it is the basic building block
for the calculation of the trimer binding energy.
As already explained the $\Xi_{cc} \bar{K}$ interaction is given by
a contact-range potential the strength of which can be related to
that of the $D K$ interaction by means of HADS.
For the form of the potential we use Eq.~(\ref{eq:V2-final}),
where we let the cutoff float in the $R_c = 0.5-2.0\,{\rm fm}$ window
but set $R_s = 0.1\,{\rm fm}$ for the second cutoff
we use to model short-range repulsion.
We determine the coupling $C(R_c)$ from the condition of reproducing
the $D_{s0}^*(2317)$ as a $DK$  bound state with a binding energy of
$45\,{\rm MeV}$,
where the resulting potential is shown in Fig.~\ref{Fig:WTpoten}.
From this potential and HADS, we predict the $\Xi_{cc} \bar{K}$ dimer to have
a binding energy of $49-64\,{\rm MeV}$ in the absence of a repulsive core,
and $49-63$ MeV if there is a repulsive core
with $C_S = 1000\,{\rm MeV}$, from which we deduce that the influence of
a repulsive core is negligible.
A more detailed compilation of the binding energies of the dimer for different
cutoffs can be found in  Table.\ref{Results:XiccXiccKbar}.~\footnote{The Coulomb interaction
is found to affect the binding energy only by several MeV, and therefore has been neglected in this work.}
We note that the binding energy of the  $\Xi_{cc}\bar{K}$  bound state is
larger than that of the $DK$ bound state because the $\Xi_{cc}$ baryon
is heavier than the $D$ meson.
%

\begin{table}[!h]
	\centering
	\caption{Parameters ($C_S$, $C(R_c)$ in MeV, and $R_s$, $R_c$ in fm) and  the binding energy ($B_2$, $B_3$ in MeV) of the $\Xi_{cc}\bar{K}$ and $\Xi_{cc}\Xi_{cc}\bar{K}$ bound states. The uncertainties are generated by
		considering a $25\%$ breaking in  HADS. }\label{Results:XiccXiccKbar}
	\begin{tabular}{c c c c}
		\hline
		\hline
		$C(R_c)$ & $R_c$ &$B_2(\Xi_{cc}\bar{K})$& $B_3(\Xi_{cc}\Xi_{cc}\bar{K})$ \\
		\hline
		&$C_S=0$ & $R_s=0.1$&\\
		\hline
		$-862.4$ & $0.5$ &$63.6^{+72}_{-51}$& $118.4^{+109}_{-78}$ \\
		$-320.1$ & $1.0$ &$53.7^{+37}_{-31}$& $92.8^{+55}_{-47}$ \\
		$-198.7$ & $1.5$ &$50.6^{+27}_{-24}$& $84.1^{+41}_{-36}, 54.2^{+31}_{-26}$ \\
		$-149.1$ & $2.0$ &$49.1^{+23}_{-21}$& $79.6^{+34}_{-31}, 55.5^{+28}_{-24}$ \\
		\hline
		&$C_S=1000$ & $R_s=0.1$&\\
		\hline
		$-884.7$ & $0.5$ &$63.2^{+70}_{-50}$& $117.5^{+106}_{-77}$ \\
		$-324.0$ & $1.0$ &$53.6^{+37}_{-31}$& $92.5^{+55}_{-41}$ \\
		$-200.2$ & $1.5$ &$50.6^{+27}_{-24}$& $83.9^{+41}_{-36}, 54.2^{+31}_{-25}$ \\
		$-149.9$ & $2.0$ &$49.1^{+23}_{-21}$& $79.6^{+34}_{-31}, 55.5^{+28}_{-24}$ \\
		\hline
	\end{tabular}
\end{table}

Now for the $\Xi_{cc}\Xi_{cc}\bar{K}$ system, besides the $\Xi_{cc}\bar{K}$
interaction explained in the previous paragraph,
we also need the $\Xi_{cc}\Xi_{cc}$ potential.
For this we use the OBE potential described
in Eqs.~(\ref{eq:V-OBE-schema}-\ref{eq:V-OBE-omega}).
With this we arrive at the results we show in Table.\ref{Results:XiccXiccKbar}.

A few comments are in order at this point.
First, the 3-body binding energy of the $\Xi_{cc}\Xi_{cc}\bar{K}$ system
increases as the cutoff $R_c$ decreases.
The origin of this cutoff dependence lies in the fact that
the $\Xi_{cc}\Xi_{cc}\bar{K}$ system is susceptible to Thomas collapse,
i.e. if we were to reduce the cutoff $R_c$ to zero, the binding energy of
the system will diverge: $B_3 \to \infty$.
For the cutoff range we have chosen, i.e. $R_c = 0.5-2.0\,{\rm fm}$,
the binding energy of the trimer varies between $80-118$ MeV.
Indeed it can be appreciated that the possibility of Thomas collapse
in the $\Xi_{cc}\Xi_{cc}\bar{K}$ system translates into a considerable
cutoff dependence of the results, yet the conclusion that the system
binds is solid, as it happens even for really soft cutoffs like
$R_c = 2.0\,{\rm fm}$.
Second, if $R_c$ is large enough ($1.5-2.0\,{\rm fm}$),
a second bound state appears.
This additional bound state is expected to be a cutoff artifact: the cutoffs
for which it appears are relatively soft, definitely larger than
the size of the hadrons we are considering here.
Be it as it may, the bottom-line is that the existence of
the $\Xi_{cc}\Xi_{cc}\bar{K}$ bound state is rather robust.


The uncertainties  of our predictions as listed
in Table.\ref{Results:XiccXiccKbar} come from two different sources.
One is violation of HADS, which is not a perfectly preserved symmetry,
but instead it is expected to be broken at the $25\%$ level.
This will affect the two two-body potentials on which the calculation of the
$\Xi_{cc} \Xi_{cc} \bar{K}$ trimer relies: the $\Xi_{cc}\bar{K}$ and
the $\Xi_{cc}\Xi_{cc}$ potentials.
In both cases we are inferring the strength of the potential from HADS and
the corresponding potential for the $D K$ and $D D$ systems,
which means that the mentioned $25\%$ relative
uncertainty applies.
In addition the individual couplings of the $\Xi_{cc}\Xi_{cc}$ potential
inherit the same type of uncertainties as in the $D D$ potential,
as discussed in Ref.~\cite{Liu:2019stu}.
The origin of these uncertainties lies in the problem of determining
the value of the different coupling constants in the OBE model,
where we simply assume them to compound into a single uncertainty
of about  $30\%$ in the value of the OBE potential,
These two sources of uncertainty --- HADS and the OBE couplings --- are then
summed in quadrature (as we expect them to be independent error sources)
\begin{equation}\label{error}
\rm{Error}=\sqrt{\rm{Error}(\rm{WT})^2+\rm{Error}(\rm{OBE})^2}
\end{equation}
where $\rm{Error}(\rm{WT})$ is calculated by scaling the $\Xi_{cc}\bar{K}$
potential by a factor of $0.75-1.25$, while keeping the OBE
$\Xi_{cc}\Xi_{cc}$ potential unchanged.
Similarly, $\rm{Error}(\rm{OBE})$ is calculated by scaling the OBE
$\Xi_{cc}\Xi_{cc}$ potential with a factor of $0.7-1.3$,
while keeping the $\Xi_{cc}\bar{K}$ potential unchanged.
We find that the influence on the binding energy form the $\rm{Error}(\rm{WT})$
is about a few tens of MeV and that from $\rm{Error}(\rm{OBE})$ is less
than 1 MeV.
Therefore, we conclude that the dominant uncertainty in the binding energy
of the trimer is the $\Xi_{cc}\bar{K}$ interaction.

\subsection{The $NN\bar{K}$ system as an analogue of
  the $\Xi_{cc}\Xi_{cc}\bar{K}$ system}

As previously mentioned, the charmed meson $D$ and
doubly charmed baryon $\Xi_{cc}$ can be seen as an analogue to the nucleon:
while the heavy quarks act as spectators, the light-quark within these
heavy hadrons belong to the same spin and isospin irreducible
representations as the nucleon (see, e.g., ~\cite{Valderrama:2019sid}).
Thus it is natural to see the $D D K$ and $\Xi_{cc}\Xi_{cc}\bar{K}$ systems
as a heavy counterpart of the $N N \bar{K}$ system.

Furthermore, the origin of the same $N\bar{K}$ interaction is the WT term
which is also responsible for the binding of the $D K$ and
$\Xi_{cc} \bar{K}$ systems.
Here there is a difference though: the nucleon belongs to the $8$
representation of the $SU(3)$-flavor group,
while the $D$ and $\Xi_{cc}$ heavy hadrons to the $\bar{3}$ and $3$ ones.
This means that the strength of the WT term is not the same for $N \bar{K}$
as it is for $D K$ and $\Xi_{cc} \bar{K}$ (in fact, for $N\bar{K}$
it is more attractive).
Using the WT term or the chiral potential as kernel to the Lippmann-Schwinger
or Bethe-Salpeter equations, one can describe the $\Lambda(1405)$ as
a $\bar{K}N$ bound state~\cite{Oset:1997it,Kaiser:1995eg,Oller:2000fj,Lutz:2001yb,Jido:2003cb,Borasoy:2004kk,Hyodo:2007jq,Hyodo:2011ur}.
In addition, the $NN\bar{K}$ system has been studied rather extensively and
it seems that all the approaches lead to the conclusion that
it should bind~\cite{Shevchenko:2006xy,Shevchenko:2007ke,Ikeda:2007nz,Yamazaki:2007cs,Arai:2007qj,Nishikawa:2007ex,Dote:2008in,Dote:2008hw,Ikeda:2008ub,Wycech:2008wf,Ikeda:2010tk,Uchino:2011jt,Barnea:2012qa,Dote:2014via,Ohnishi:2017uni,Dote:2017veg}, while only differing in minor details.
If this were not enough,
all the experiments performed so far support the existence of
such a state\cite{Agnello:2005qj,Yamazaki:2008hm,Yamazaki:2010mu,Ichikawa:2014ydh,Sada:2016nkb,Ajimura:2018iyx}, again only differing in minor details.

Following the logic with which we studied the $\Xi_{cc}\Xi_{cc}\bar{K}$ system,
here we calculate the binding energy of the $NN\bar{K}$ bound state
using the $N\bar{K}$ and $NN$ potentials as input.
For fixing the strength of the $N\bar{K}$ interaction we simply reproduce
the location of the $\Lambda(1405)$ with the potential of
Eq.~(\ref{eq:V2-final}).
For the $NN$ interaction we use the OBE model.
The binding energy of the $NN\bar{K}$ trimer is listed
in Table \ref{Results:NNKbar} for different cutoffs.
In the same table we also show the values of the coupling that reproduces
the $\Lambda(1405)$ pole as a $N\bar{K}$ bound state with a binding energy
$B_2 = 29.4\,{\rm MeV}$.
The binding energy of the $NN\bar{K}$ trimer ranges from $35-43\,{\rm MeV}$,
where the cutoff dependence is relatively weak in comparison to
the $\Xi_{cc} \Xi_{cc} \bar{K}$ system.
The reason is that the $NN\bar{K}$ does not suffer from Thomas collapse
as a consequence that the mass ratio between the nucleon and the kaon
is not large enough as to trigger this effect.
We notice that the inclusion of the $NN\bar{K}$ system in the present
manuscript should be viewed mostly as a consistency check of
the $\Xi_{cc} \Xi_{cc} \bar{K}$ calculation: there is a large
literature of calculations of the $NN\bar{K}$ system that
are far more sophisticated that the one presented here.

\begin{table}[!h]
	\centering
	\caption{Parameters ($C_S$, $C(R_c)$ in MeV, and $R_s$, $R_c$ in fm) of the $N\bar{K}$ potential and the binding energies ($B_2$ and $B_3$ in MeV) of the $N\bar{K}$ and $I(J^P)$=$\frac{1}{2}(0^-)$ $NN\bar{K}$ bound states. The parameters are determined by reproducing the $\Lambda(1405)$ with the binding energy 29.4 MeV with respect to the $N\bar{K}$ threshold.}\label{Results:NNKbar}
	\begin{tabular}{ c c c c}
		\hline
		\hline
		$C(R_c)$& $R_c$ & $B_2(N\bar{K})$ & $B_3(NN\bar{K})$\\
		\hline
		&$C_S=0$& $R_s=0.1$&\\
		\hline
		$-925.9$ & $0.5$ & $29.4$ & $35.2$ \\
		$-316.4$ & $1.0$ & $29.4$ & $39.3$ \\
		$-132.6$ & $2.0$ & $29.4$ & $41.8$ \\
		$-89.2$ & $3.0$ & $29.4$ & $42.5$ \\
		\hline
		&$C_S=1000$& $R_s=0.1$&\\
		\hline
		$-946.6$ & $0.5$ & $29.4$ & $35.4$ \\
		$-319.8$ & $1.0$ & $29.4$ & $39.4$ \\
		$-133.2$ & $2.0$ & $29.4$ & $41.8$ \\
		$-89.4$ & $3.0$ & $29.4$ & $42.5$ \\
		\hline
	\end{tabular}
\end{table}

\section{Summary}
\label{sec:Summary}

In this work we have investigated the $\Xi_{cc} \Xi_{cc} \bar{K}$ system.
We have reached the conclusion that it likely binds.
This is mostly a consequence of HADS, a type of heavy-quark symmetry that
relates the $\Xi_{cc} \bar{K}$ interaction to the $D K$ one.
From this symmetry and the fact that previous theoretical explorations point out
to the possibility of bound $D D K$~\cite{SanchezSanchez:2017xtl,MartinezTorres:2018zbl}
and $D D D K$ clusters~\cite{Wu:2019vsy},
the natural expectation is that the $\Xi_{cc} \Xi_{cc} \bar{K}$
system binds too.
Concrete calculations within the GEM framework indicate that
(i) the two-body $\Xi_{cc}\bar{K}$ system has a binding energy
$B_2 = 49-64\,{\rm MeV}$ (ii) while for the three-body
$\Xi_{cc} \Xi_{cc} \bar{K}$ system the binding energy is
$B_3 = 80-118\,{\rm MeV}$.

In the case of the two-body calculation, from HADS we expect
the $\Xi_{cc} \bar{K}$ potential to be identical to the $D K$ one,
where for the later case the strength of the potential can be completely
determined from the hypothesis that the $D_{s0}^*(2317)$ is a $D K$ bound state.
At the level of approximation we are considering, this interaction is given by
the WT term, though we additionally considered the possibility of a short-range
repulsive core in the $\Xi_{cc} \bar{K}$ potential to mimic the effect of
subleading corrections.
The uncertainty in the two-body binding energy comes mostly from violations of
HADS, which we take to be as large as  25\% percent.

For the three-body $\Xi_{cc}\Xi_{cc}\bar{K}$ trimer the uncertainties are
definitely larger because besides HADS we also have a sizable
cutoff dependence: this trimer is in principle susceptible to Thomas collapse,
i.e. if the range of the $\Xi_{cc} \bar{K}$ interaction were to be taken to zero
the trimer binding energy would diverge, hence the cutoff dependence.
This problem can be effectively circumvented by the fact that the size of
the $\Xi_{cc}$ or $\bar K$ hadrons is finite or, more elegantly,
by the inclusion of a repulsive short-range three-body force
that will stabilize the results.
Here we opt for the finite-cutoff solution, owing to its simplicity but also
to the fact that the strength of a prospective three-body force should be
determined from the data.
Be it as it may, we consider the existence of a relatively compact
$\Xi_{cc}\Xi_{cc}\bar{K}$ trimer as a robust conclusion.
Besides, we can deduce the existence of additional trimers from the other
heavy-quark symmetries.
For instance, from heavy-flavor symmetry there should be
$\Xi_{bc} \Xi_{bc} \bar{K}$ and $\Xi_{bb} \Xi_{bb} \bar{K}$ trimers.

Finally we applied our framework to study
the $N\bar{K}$ and $NN\bar{K}$ systems.
The motivation is a very simple analogy between the charmed mesons $D$,
the doubly charmed baryons $\Xi_{cc}$ and the nucleon $N$, the three of
which belong to the same representation of light-quark spin
and isospin.
From this analogy the only differences between few nucleons and
few doubly-charmed baryons systems are the specific details of the potential,
i.e. the coupling constants and masses.
If we determine the $N\bar{K}$ interaction from the condition that
the $\Lambda(1405)$ is a $N \bar{K}$ bound state,
we reach the conclusion that the binding energy of the $NN\bar{K}$ system
is $B_3 = 35-43\,{\rm MeV}$ with respect to the three-body mass threshold,
which is consistent with previous studies and experiments.

\section{Acknowledgements}
This work is partly supported by the National Natural Science Foundation of China under Grant Nos.11735003, 11975041,  and 11961141004, the fundamental Research Funds for the Central Universities, and the Thousand Talents Plan for Young Professionals.

\bibliography{Xicc-Xicc-Kbar}

\end{document}